\documentclass[prd,nofootinbib,showpacs,showkeys,draft]{revtex4}
\usepackage{epic}
\def\mb#1         {\mbox{\boldmath $#1$}}

\begin{document}
%
%
\title{$0^+$ and $1^+$ States of $B$ and $B_s$ Mesons}
%
%
\author{Takayuki Matsuki}
\email[E-mail: ]{matsuki@tokyo-kasei.ac.jp}
\affiliation{Tokyo Kasei University,
1-18-1 Kaga, Itabashi, Tokyo 173, JAPAN}
\author{Kentarou Mawatari}
\email[E-mail: ]{mawatari@radix.h.kobe-u.ac.jp}
\author{Toshiyuki Morii}
\email[E-mail: ]{morii@kobe-u.ac.jp}
\affiliation{Graduate School of Science and Technology,
Kobe University,\\ Nada, Kobe 657-8501, JAPAN}
\author{Kazutaka Sudoh}
\email[E-mail: ]{sudou@rarfaxp.riken.go.jp}
\affiliation{Radiation Laboratory, RIKEN (The Institute of Physical and Chemical Research), \\ 
Wako, Saitama 351-0198, JAPAN}

\date{November 10, 2004}

\begin{abstract}
Predicted masses of $J^P=0^+$ and $1^+$ states of $D_{sJ}$ and $D$ by a
potential model proposed some time ago by two of us (T.M. and T.M.),
in which the Hamiltonian and wave functions are expanded in $1/m_Q$ 
with $m_Q$ heavy quark mass respecting heavy quark symmetry, have 
recently been confirmed by BaBar and Belle experiments within one 
percent accuracy. 

In this Letter, decay modes of $0^+$ and $1^+$ states of $B$ and $B_s$ mesons
are discussed using the predicted masses of this model.
\end{abstract}
\preprint{}
\maketitle

%
\section{Introduction}
\label{intro}
%

The BaBar's discovery \cite{BaBar} of $D_{sJ}(2317)$ and $D_{sJ}(2460)$ of the
system $c$ and $s$ quarks has inspired theorists \cite{BEH,BH} to explain
these states in some way because a well-known potential model \cite{GI,GK}
failed to expound masses of these states. This discovery has been soon
confirmed by CLEO and Belle \cite{CLEO}. These $D_{sJ}$ states are narrow
because isospin violation and $DK$/$D^*K$ threshold prohibit them to decay
into lower states $D_s(1968)$ and $D_s^*(2112)$. This is another reason why
theorists are interested in these states and are trying to explain their
branching ratios, too.

Soon after the discovery, another set of heavy mesons, $D_0^{*0}(2308)$ and
$D_1^{'0}(2427)$ of $c$ and $u/d$ quarks which have the same quantum numbers
$J^P=0^+$ and $1^+$ as $D_{sJ}$, has been discovered by Belle \cite{ABE}.
Again masses of these states cannot be fitted with those of a potential model
\cite{GI,GK}. These states $D_0^{*0}$ and $D_1^{'0}$ have no restriction
like $D_{sJ}$ and hence their decay (like $D\pi$ and $D^*\pi$) width becomes
broad.

To explain masses of $D_{sJ}$, Bardeen, Eichten, Hill, and others \cite{BEH,BH}
proposed an interesting idea of an effective Lagrangian with chiral symmetries
of light quarks and heavy quark symmetry. The heavy meson states with the
total angular momentum $J=0$ and $J=1$ with $j_q=1/2$, which is the total
angular momentum of the light quark degrees of freedom, make the parity
doublets $(0^-, 0^+)$ and $(1^-, 1^+)$, respectively, and the members in each
doublet degenerate in a limit of chiral symmetry. Furthermore, the two states
$(0^-, 1^-)$ degenerate in a limit of heavy quark symmetry, as well as
$(0^+, 1^+)$ do. These doublets, $(0^-,1^-)$ and $(0^+,1^+)$, are called the
heavy spin multiplets. Though their model can reproduce well the 
mass difference between members in each parity doublet
by using the modified Goldberger-Treiman relation, unfortunately it
cannot calculate masses themselves.  This is why
some people have proposed a modified potential model by using a $DK$ bound
state \cite{BCL}.

Several years ago, Godfrey, Isgur, and Kokoski proposed a 
relativised potential model \cite{GI,GK},
in which they included $\sqrt{p^2+m^2}$ as a kinetic term and 
spin-dependent interaction terms where linear-rising confining as well as
short-range Coulomb potentials are taken into account. This model, however,
could not reproduce the masses of $D_{sJ}(2317)$ and $D_{sJ}(2460)$. 
Even though they have taken an infinite heavy quark mass limit in 
the second paper \cite{GK}, they could not reproduce $D_{sJ}(1^+)$ mass
and their calculated masses exceeded $DK/D^*K$ threshold.
Thus, there appear some discussions \cite{BCL} that a potential model 
is not appropriate to describe these states. 

The above potential model does not completely and consistently respect heavy
quark symmetry and does not treat quarks as Dirac particles. Some time ago,
two of the authors (T.M. and T.M.) \cite{MM} proposed a new bound state
equation for atomlike mesons, i.e., heavy mesons composed of a heavy quark and
a light antiquark.  In this model compared
with \cite{GI,GK}, quarks are treated as four-spinor particles from the
beginning and both the Hamiltonian and wave functions are expanded in $1/m_Q$
so that our model treats quarks as relativistic as possible and consistently
takes into account heavy quark symmetry within a potential model. Our predicted
masses for these states, $D_{sJ}$ and $D_0^{*0}$ and $D_1^{'0}$, are in good
agreement within one percent accuracy when calculated at the first order in
$1/m_Q$. See Table \ref{table1} in this letter and Tables III and IV
in \cite{MM}.
\begin{table*}[t]
\caption{Comparison of higher $D$ and $D_s$ meson masses (units in MeV)}
\label{table1}
\begin{tabular*}{10cm}{c|@{\extracolsep{\fill}}cccc}
\hline
\hline
  \makebox[1.5cm]
  {$J^P$} & $D(0^+)$ & $D(1^+)$ & $D_s(0^+)$ & $D_s(1^+)$ \\
\hline
  observed  & 2308 & 2427 & 2317 & 2457 \\
  predicted  & 2304 & 2449 & 2339 & 2487 \\
\hline
\end{tabular*}
\end{table*}

In this Letter, having confirmed that our model has well succeeded in
predicting masses of recently discovered heavy mesons, we predict masses 
of $0^+$ and $1^+$ of $B$ and $0^+$, $1^-$ and $1^+$ of $B_s$ mesons 
by citing the evaluated values in \cite{MM}. Here we also discuss 
their decay modes; whether
these mesons violate isospin symmetry or not and whether mass difference
between $0^+$ ($1^+$) and $0^-$ ($1^-$) is less than the $BK$/$B^*K$-mass
threshold or not. We expect these higher states of $B$ and $B_s$ mesons can
be detected in CDF/LHC experiments.

%
\section{Higher $B$ and $B_s$ meson masses and their Decay Modes}
\label{Bmass}
%
Our prediction in \cite{MM} of $B$ and $B_s$ meson masses in the first order
of $1/m_Q$ corrections is given in Table \ref{table2}.
\begin{table*}[t]
\caption{Higher $B$ and $B_s$ meson masses taken from \cite{MM} (units in MeV)}
\label{table2}
\begin{tabular*}{10cm}{c|@{\extracolsep{\fill}}ccccc}
\hline
\hline
  \makebox[1.5cm]
  {$J^P$} & $B(0^+)$ & $B(1^+)$ & $B_s(1^-)$ & $B_s(0^+)$ & $B_s(1^+)$ \\
\hline
  predicted  & 5697 & 5740 & 5440 & 5716 & 5760 \\
\hline
\end{tabular*}
\end{table*}
Considering the fact that $D$/$D_s$ masses have been well
predicted by our model within one percent accuracy as seen in Table
\ref{table1}, we expect masses of these $B$ and $B_s$ mesons are similarly
within one percent accuracy to be observed. Furthermore when looking at
Table \ref{table1} carefully, we expect that our predicted masses of $B$/$B_s$
mesons agree with experiments better than those of $D_s$ since $m_{u,d}/m_b <
m_{u,d}/m_c <  m_s/m_b < m_s/m_c \ll 1$ ($m_{u,d}=10$ MeV, $m_s=94.72$ MeV,
$m_c=1457$ MeV, and $m_b=4870$ MeV are adopted in our paper \cite{MM}).

Let us discuss decay modes of $0^+$ and $1^+$ states of $B$ and $B_s$ mesons
taking into account these meson masses given by Table \ref{table2}.
\def\labelenumi{(\theenumi)}
\begin{enumerate}
\item $B_s(1^-) \rightarrow B_s(0^-) + \gamma$
\item $B(0^+) \rightarrow B(0^-) + \pi$ \quad{\rm ~with~broad~decay~width}
\item $B(1^+) \rightarrow B(1^-) + \pi$ \quad{\rm ~with~broad~decay~width}
\item $B_s(0^+) \rightarrow B_s(0^-) + \pi$ \quad{\rm ~with~narrow~decay~width}
\item $B_s(1^+) \rightarrow B_s(1^-) + \pi$ \quad{\rm ~with~narrow~decay~width}
\end{enumerate}
Comments are given as follows.
\def\labelenumi{(\theenumi)}
\begin{enumerate}
\item Decay $B_s(1^-) \rightarrow B_s(0^-) + \gamma$ is
      similar to $B(1^-) \rightarrow B(0^-) + \gamma$,
      i.e. $B^* \rightarrow B + \gamma$, and is dominant decay mode.
\item Decay width of $B(0^+) \rightarrow B(0^-) + \pi$ is as broad as a few
      hundred MeV like $D(0^+) \rightarrow D(0^-) + \pi$ \cite{ABE} because
      this is strong decay and is not prohibited by isospin invariance since
      $I(B(0^-))=I(B(0^+))=1/2$ while $I(\pi)=1$, where $I(X)$ is isospin of
      a particle $X$.
\item Decay width of $B(1^+) \rightarrow B(1^-) + \pi$ is also as broad as a
      few hundred MeV like $D(1^+) \rightarrow D(1^-) + \pi$ \cite{ABE} because
      this is strong decay and is not prohibited by isospin invariance since
      $I(B(1^-))=I(B(1^+))=1/2$ while $I(\pi)=1$.
\item Decay width of $B_s(0^+) \rightarrow B_s(0^-) + \pi$ is expected very
      narrow, a few MeV like the decay of $D_s(0^+) \rightarrow D_s(0^-) +
      \pi$ since this decay mode is prohibited by isospin invariance due to
      $I(B_s(0^+))=I(B_s(0^-))=0$ while $I(\pi)=1$ and the predicted mass of
      $B_s(0^+)$ is below $BK$ threshold.
\item Decay width of $B_s(1^+) \rightarrow B_s(1^-) + \pi$ is also expected
      very narrow, a few MeV like the decay of $D_s(1^+) \rightarrow D_s(1^-)
      + \pi$ since this decay mode is prohibited by isospin invariance due to
      $I(B_s(1^+))=I(B_s(1^-))=0$ while $I(\pi)=1$ and the predicted mass of
      $B_s(1^+)$ is below $B^*K$ threshold.
\end{enumerate}
Here $B_s(0^-)=B_s(5370)$, $D_s(0^-)=D_s^\pm(1968)$,
$D_s(1^-)=D_s^{*\pm}(2112)$, $D_s(0^+)=D_{sJ}(2317)$, and
$D_s(1^+)=D_{sJ}(2460)$ \cite{PP}. We expect these higher states of $B$ and
$B_s$ mesons can be detected in CDF/LHC experiments by looking at their decay
modes.

%
\section{Chiral Limit of $H_0$}
\label{chiralH0}
%

As mentioned above, the model by Bardeen and Hill \cite{BH} has
chiral symmetry in the chiral limit of light quark mass,
$m_q\rightarrow 0$.  Our model has also chiral symmetry.  
Here let us see how it is realized.

In our model the lowest order mass of the $Q\bar q$ bound state 
is given by $m_Q+E_0^\ell$
after solving the following Schr\"odinger equation \cite{MM,MMMS}, 
\begin{equation}
  H_0 \otimes\psi _0^\ell = E_0^\ell\psi _0^\ell, \qquad
  H_0 =
  \vec\alpha_q\cdot\vec p+\beta_q\left(m_q+S(r)\right)+V(r), \label{h0}
\end{equation}
where $H_0$ is the lowest order Hamiltonian, 
$\ell$ expresses all the quantum numbers, $\psi_0^\ell$ is the
lowest solution to Eq. (\ref{h0}), and with a notation $\otimes$ one should
understand that gamma
matrices for a light antiquark be multiplied from left with the wave function
while those for a heavy quark from right. Here $S(r)$ is a confining scalar
potential and $V(r)$ is a Coulombic vector potential at short distances. Both
potentials have dependency only on $r$, relative distance between $Q$ and
$\bar q$. Quantities with a subscript $q$ mean those for a light antiquark.
Using the spherical polynomials $Y_j^m$ and the vector spherical harmonics
defined by
\begin{eqnarray}
  \vec Y_{j\,m}^{(\rm L)} = -\vec n\,Y_j^m, \quad
  \vec Y_{j\,m}^{(\rm E)} = {r \over {\sqrt {j(j+1)}}}\vec \nabla Y_j^m,
  \quad
  \vec Y_{j\,m}^{(\rm M)} = -i\vec n\times \vec Y_{j\,m}^{(\rm E)},
\end{eqnarray}
we can decompose the wave function $\psi_0^\ell$ into radial and angular
parts as follows \cite{MM,MMMS};
\begin{equation}
  \psi _0^\ell = \left( {\matrix{ 0  & \Psi _{j\,m}^k(\vec r) }} \right),
  \label{0thsols1}
  \qquad
  \Psi _{j\,m}^k(\vec r) = \frac{1}{r}
  \left( {\matrix{f_k(r)\;y_{j\,m}^k\cr ig_k(r)\;y_{j\,m}^{-k}\cr}} \right),
  \label{0thsols2}
\end{equation}
where the angular part $y_{j\,m}^{\pm k}$ is given by the 
linear combination of $Y_j^m$ and
$\vec\sigma\cdot\vec Y_{j\,m}^{(\rm A)}$ (A=L, M, E).
Substituting this wave function into Eq. (1), one can obtain the radial 
part equation as follows; 
\begin{equation}
  \left( {\matrix{{m_q+S+V}&{-\partial _r+{k \over r}}\cr
  {\partial _r+{k \over r}}&{-m_q-S+V}\cr
  }} \right) \Psi _k(r) = E^k_0\;\Psi _k(r), 
  \qquad
  \Psi _k(r)\equiv\left( {\matrix{{f_k\left( r \right)}\cr
  {g_k\left( r \right)}\cr}} \right). \label{radial0}
\end{equation}
Here $k$ is the quantum number of the spinor operator $K$ 
defined by \cite{MM,MMMS}
\begin{equation}
  K = -\beta_q \left( \vec \Sigma_q \cdot \vec L + 1 \right),
  \qquad
  K\, \Psi_{j\,m}^k = k\, \Psi_{j\,m}^k.
  \label{k_quantum}
\end{equation}
where $\vec\Sigma_q\ (=\vec\sigma_q\;1_{2\times 2})$ and $\vec L$ are
the 4-component spin and the orbital angular momentum of the light antiquark,
respectively.  Note that with this quantum number $k$, one can simultaneously
determine both a partial angular momentum $j_q$ of light degrees 
of freedom and a total parity $P$ as \cite{MMMS},
\begin{equation}
  j_q = |k| - \frac{1}{2},
  \qquad
  P = \frac{k}{|k|} (-)^{|k| + 1}.
\end{equation}
It is remarkable that in our approach $K$
can be defined even for a two-body bound system
composed of a heavy quark and a light antiquark.

Then, in our model a chiral limit is realized by setting 
$m_q=S(r) =0$, in which
the corresponding Hamiltonian becomes
\begin{equation}
  H_0^{chiral}=\vec \alpha_q\cdot\vec p + V(r). \label{chiralH}
\end{equation}
With this chiral Hamiltonian, the radial part equation becomes
\begin{equation}
  \left( {\matrix{{V}&{-\partial _r+{k \over r}}\cr
  {\partial _r+{k \over r}}&{V}\cr
  }} \right) \Psi _k^{chiral}(r) = 
  E^{k~(chiral)}_0\;\Psi _k^{chiral}(r),
  \label{radial1}
\end{equation}
This equation can be converted into the one with $-k$ by a unitary
transformation with $U=\sigma_2$, where $\sigma_2$ denotes the 2nd 
component of Pauli matrices,
\begin{equation}
  \left( {\matrix{{V}&{-\partial _r-{k \over r}}\cr
  {\partial _r-{k \over r}}&{V}\cr
  }} \right) U \Psi _k^{chiral}(r) = 
  E^{k~(chiral)}_0\;U \Psi _k^{chiral}(r), \label{radial2}
\end{equation}
which means,
\begin{equation}
  U \Psi _k^{chiral}(r) = \Psi _{-k}^{chiral}(r) \quad
  {\rm ~and~}\quad E^{k~(chiral)}_0 = E^{{-k}~(chiral)}_0.
\end{equation}
That is, the energy is degenerate in $\pm k$ in the case of chiral Hamiltonian,
Eq. (\ref{chiralH}), which corresponds to the degeneracy for $k=\pm 1$ 
in the chiral limit. The degenerate hadron mass is given by $m_Q$ since
Eq.(\ref{radial1}) is nothing but an equation for a hydrogen atom with
mass $=0$ and its energy level is proportional to mass which is zero in this
case, i.e., $E_0^{k~(chiral)}=0$.

Considering the above observation,
mass splitting in our model occurs as follows.
\begin{enumerate}
\item Start from the chiral limit Hamiltonian, Eq. (\ref{chiralH}),
      together with no $1/m_Q$ corrections with $m_Q$ heavy quark mass.
      In this stage, all the masses of $0^{-}$, $0^{+}$, $1^{-}$ and
      $1^{+}$ states with $j_q=1/2$ are degenerate;
      $m_Q=m(0^-)=m(1^-)=m(0^+)=m(1^+)$.
\item When the light quark mass $m_q$ and a scalar potential $S(r)$, i.e. 
      explicit chiral breaking terms are inserted as shown in Eq. (\ref{h0}), 
      degeneracy is partially broken; $m(0^-)=m(1^-)$ and $m(0^+)=m(1^+)$, 
      which are called heavy quark
      multiplets in \cite{BEH}, because there still remains degeneracy due to
      the quantum number $k$ \cite{MMMS}.
\item Finally by including $1/m_Q$ terms, all the degeneracy is resolved
      and the mass values in Tables \ref{table1} and \ref{table2} are given.
\end{enumerate}

Notice that the quantum number $k$ plays an important role in our model
to see how the states are classified and how the degeneracy is
resolved \cite{MMMS}. 

%
%
%

The above procedure may be depicted in Figure 1.
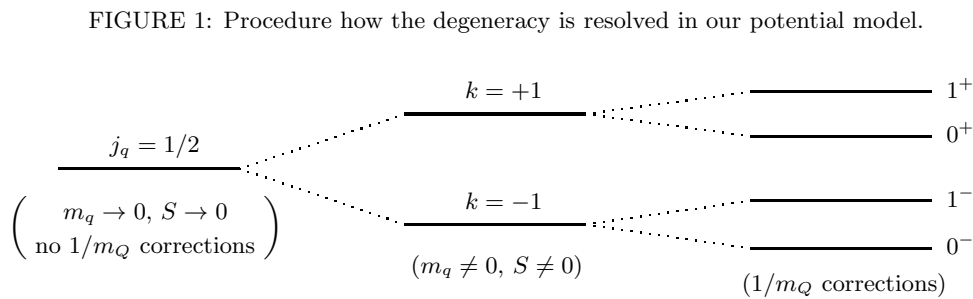
\begin{center}
\begin{figure}[htb]
\label{figure1}
\caption{Procedure how the degeneracy is resolved in our potential model.}
\begin{picture}(350,90)
\setlength{\unitlength}{0.4mm}
 \thicklines
  \put(  5,36.5){\line(1,0){60}}
  \put(120,54.7){\line(1,0){60}} \put(120,17.95){\line(1,0){60}}
  \put(235,62.1){\line(1,0){60}} \put(235,47.3 ){\line(1,0){60}}
  \put(235,25.9){\line(1,0){60}} \put(235,10.0 ){\line(1,0){60}}
 \thinlines
  \dottedline{3}( 65,36.5 )(120,54.7) \dottedline{3}( 65,36.5 )(120,17.95)
  \dottedline{3}(180,54.7 )(235,62.1) \dottedline{3}(180,54.7 )(235,47.3 )
  \dottedline{3}(180,17.95)(235,25.9) \dottedline{3}(180,17.95)(235,10.0 )
  \put(22,42){$j_q=1/2$}
  \put(-12,15){$\left(\begin{array}{c}
              m_q\to0,\, S\to0 \\  {\rm no}~ 1/m_Q~ {\rm corrections}
             \end{array}\right)$}
  \put(140,60){$k=+1$}        \put(140,23){$k=-1$}
  \put(300,60.1){$1^+$} \put(300,45.3){$0^+$}
  \put(300,23.9){$1^-$} \put(300, 8.0){$0^-$}
 \put(122,2){($m_q\neq 0,\, S\neq0$)}    \put(232,-4){($1/m_Q$ corrections)}
\end{picture}
\end{figure}
\end{center}

On the other hand, the procedure taken by Bardeen and Hill \cite{BH}
can be stated as follows.
\begin{enumerate}
\item Start from the chiral limit of light quark masses but with 
      infinite heavy
      quark mass instead of finite.  In this stage, all the masses 
      of $0^{-}$, $0^{+}$, $1^{-}$, and $1^{+}$ states with $j_q=1/2$
      are degenerate just like our model.  
\item Inclusion of a finite heavy quark mass leads to a partial
      resolution of degeneracy;
      $m(0^-)=m(0^+)$ and $m(1^-)=m(1^+)$, which are called parity doublets.
\item Finally by including a finite light quark mass effects and using 
      the modified Goldberger-Treiman relation, all the degeneracy 
      is resolved.
\end{enumerate}
The above procedure may be depicted in Figure 2.
\begin{center}
\begin{figure}[htb]
\label{figure2}
\caption{Procedure how the degeneracy is resolved in \cite{BEH}.}
\begin{picture}(350,85)
\setlength{\unitlength}{0.4mm}
 \thicklines
  \put(  5,34.55){\line(1,0){60}}
  \put(120,44.13){\line(1,0){60}} \put(120,27.4){\line(1,0){60}}
  \put(235,59.1 ){\line(1,0){60}} \put(235,44.8){\line(1,0){60}}
  \put(235,24.3 ){\line(1,0){60}} \put(235,10.0){\line(1,0){60}}
 \thinlines
  \dottedline{3}( 65,34.55)(120,44.13) \dottedline{3}( 65,34.55)(120,27.4)
  \dottedline{3}(180,44.13)(235,59.1 ) \dottedline{3}(180,44.13)(235,24.3)
  \dottedline{3}(180,27.4 )(235,44.8 ) \dottedline{3}(180,27.4 )(235,10.0)
  \put(22,41){$j_q=1/2$}
  \put(11,15){$\left(\begin{array}{c}
              m_q\to0 \\ m_c\to\infty 
             \end{array}\right)$}
  \put(140,48){$J^P=1^\pm$}     \put(140,30.5){$J^P=0^\pm$}
  \put(300,57.1){$1^+$} \put(300,42.8){$0^+$}
  \put(300,22.3){$1^-$} \put(300, 8.0){$0^-$}
  \put(251,-4){($m_q\neq 0$)}    \put(131,12){($m_c$ finite)}
\end{picture}
\end{figure}
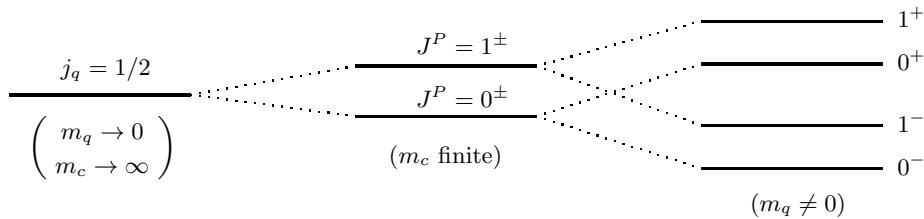
\end{center}

Comparing these two procedures, one can easily see that our model
naturally explains how to resolve degeneracy between $0^-$ and $1^-$
and/or $0^+$ and $1^+$ due to the quantum number $k$ and clarify
the origin of mass splitting. Namely the interpretation by Bardeen, Eichten,
and Hill is not the unique way to explain the mass splitting. There
still remains a way to explain it using a potential model. Moreover,
it should be noted that the potential model can give not only the 
mass differences of heavy mesons but also their absolute 
values, as seen in Tables \ref{table1} and \ref{table2}.

%
\section{Check of the Goldberger-Treiman Relation}
\label{GDrelation}
%

Before closing this paper, we would like to give some comments
on the modified Goldberger-Treiman relation  claimed by
Bardeen, Eichten, and Hill \cite{BEH}. The modified 
Goldberger-Treiman relation predicts $\Delta M(0) = \Delta M(1)$,
where 
\[
  \Delta M(0) = M(0^+)-M(0^-), \qquad
  \Delta M(1) = M(1^+)-M(1^-).
\]
Then, one can test it by using the following mass differences 
for a couple of states.
\begin{enumerate}
\item $D_s(c\bar s)$ states \cite{BaBar}
\begin{eqnarray*}
  \Delta M(0) &=& D_{sJ}(0^+, 2317)-D_s(0^-, 1969)=348 {\rm ~MeV}
  \\
  \Delta M(1) &=& D_{sJ}(1^+, 2460)-D_s^*(1^-, 2112)=348 {\rm ~MeV}
\end{eqnarray*}
\item $D(c\bar u(\bar d))$ states \cite{ABE}
\begin{eqnarray*}
  \Delta M(0) &=& D(0^+, 2308)-D(0^-, 1870)=438 {\rm ~MeV}
  \\
  \Delta M(1) &=& D(1^+, 2420)-D(1^-, 2010)=410 {\rm ~MeV}
\end{eqnarray*}
\item $K(s\bar u(\bar d))$ states \cite{PP}
\begin{eqnarray*}
  \Delta M(0)=K(0^+, 1430)-K(0^-, 495)=935 {\rm ~MeV}
  \\
  \Delta M(1)=K(1^+, 1335)-K(1^-, 895)=440 {\rm ~MeV}
\end{eqnarray*}
\end{enumerate}
where we have given mass of $K(1^+)$ =1335 MeV as a simple average of
$K_1(1270)$ and $K_1(1400)$. $D(0^+, 2308)$ is from the Belle 
data \cite{ABE} and others are from the particle data group \cite{PP}. 
As one can see here, the modified Goldberger-Treiman relation looks to 
work amazingly well for $D_s(c\bar s)$ states.  However, 
it might be problematic for other cases as shown in the following,  
\begin{enumerate}
\item the mass of $D(0^+)$ is expected to be 2280 MeV, 
if $\Delta M(0)=\Delta M(1)=410$ MeV holds, which is quite 
different from 2308 MeV observed by Belle \cite{ABE}.   
Compared to this expectation, it is remarkable that in our case 
it is estimated to be 2304 MeV, as shown in Table I, being very close to
the Belle data. 
%
%
\item furthermore in the case of $K$ states, it seems this 
relation cannot be applied, though its application to this case 
might not be justified because $s$ quark is not heavy.
\end{enumerate}
Based on this observation, the modified Goldberger-Treiman relation
might be correct only in the
case of $D(c\bar s)$ and there might be no more meaning for other states.
To confirm this, it is very important to observe $B(0^+)$ and $B(1^+)$
as well as $B_s(0^+)$ and $B_s(1^+)$, though a collective mass of 
orbitally excited ($L=1$) $B$ meson states was reported 
a couple of years ago \cite{CDF}.  
We strongly wish that these particles will be observed soon.

\def\Journal#1#2#3#4{{#1} {\bf #2}, #3 (#4)}
\def\NIM{Nucl. Instrum. Methods}
\def\NIMA{Nucl. Instrum. Methods A}
\def\NPB{Nucl. Phys. B}
\def\PLB{Phys. Lett. B}
\def\PRL{Phys. Rev. Lett.}
\def\PRD{Phys. Rev. D}
\def\ZPC{Z. Phys. C}
\def\EPJ{Eur. Phys. J. C}
\def\PR{Phys. Rept.}
\def\IJM{Int. J. Mod. Phys. A}

\end{document}